\documentclass[letterpaper,12pt]{article}

\usepackage{graphicx}       
\usepackage{chemformula}    
\usepackage{authblk}
\usepackage[margin=1in,letterpaper]{geometry} 
\usepackage{tabularx}
\usepackage{xcolor}
\newcommand{\negSB}[1]{\textcolor{red}{\textbf{#1}}}
\usepackage[table]{xcolor} 
\usepackage[dvipsnames]{xcolor} 

\begin{document}

\title{Effect of Exchange-Correlation Functionals on Schottky Barriers at Si/Metal Interfaces}

\author[1,2]{Viviana Dovale-Farelo}
\author[1,3,4]{Kamal Choudhary}
\affil[1]{National Institute of Standards and Technology, Gaithersburg, MD, 20899}
\affil[2]{University of Maryland, College Park, MD, 20742}

\affil[3]{Department of Materials Science and Engineering, Johns Hopkins University, Baltimore, MD 21218, USA}
\affil[4]{Department of Electrical and Computer Engineering, Johns Hopkins University, Baltimore, MD 21218, USA}

\date{\today}
\maketitle

\begin{abstract}
Accurate prediction of Schottky barrier heights (SBHs) at metal-semiconductor (M-SC) interfaces is essential for understanding and optimizing charge injection in electronic and optoelectronic devices. However, first-principles calculations of SBHs remain challenging due to the combined difficulties of semiconductor bandgap underestimation, metal Fermi level placement, lattice-mismatch, relative geometric alignment and electrostatic potential alignment across heterogeneous interfaces. In this work, we present a systematic and physically grounded assessment of computational strategies for SBH prediction using Si(111)/Metal (Al, Cu, Ag, Au) interfaces as representative test cases. We evaluate multiple exchange-correlation (XC) treatments, in combination with three distinct bulk reference protocols: relaxed bulk, relaxed bulk with spin-orbit coupling, and strained bulk references consistent with the interface geometry. By benchmarking against experimental data, we demonstrate that structural and electrostatic consistency between interface and bulk reference calculations is the dominant factor governing SBH accuracy. We show that mixed hybrid-semilocal approaches combined with strained reference protocols yield uniformly positive and significantly improved SBHs, achieving near-experimental accuracy while maintaining a favorable balance between computational cost and predictive performance. Our results establish a clear and transferable methodology for reliable Schottky barrier prediction and provide practical guidance for large-scale screening and interface engineering.
\end{abstract}


\section{Introduction}
Metal-semiconductor (M-SC) interfaces play a central role in determining the performance of electronic and optoelectronic devices, including diodes, transistors, and sensors \cite{robertson2013band, milnes2012heterojunctions}. A key property governing charge transport across these interfaces is the Schottky barrier height (SBH), which controls carrier injection from the metal into the semiconductor. The SBH originates from the alignment of the metal Fermi level ($E_F$) with the electronic band structure of the semiconductor upon contact \cite{schottky1926small, bardeen1947surface}. For electrons, the barrier height $\Phi_n$ is defined with respect to the conduction band minimum (CBM), while for holes, the barrier height $\Phi_p$ is referenced to the valence band maximum (VBM) \cite{tung2014physics}. These barriers directly influence contact resistance, turn-on voltage, leakage current, and overall device efficiency.

Accurate knowledge of Schottky barriers is therefore essential for the rational design and optimization of electronic and optoelectronic components. In modern nanoscale devices, contact properties often dominate device performance, making reliable prediction of SBHs a prerequisite for materials selection and interface engineering. As device dimensions shrink and new material combinations are explored, predictive modeling of metal-semiconductor \cite{choudhary2025slakonet} contacts becomes increasingly important.

Despite their conceptual simplicity, Schottky barriers remain challenging to predict quantitatively from first principles. Standard density functional theory (DFT) \cite{1965dft} approaches are limited by the well-known bandgap underestimation problem associated with commonly used exchange-correlation (XC) functionals such as the local density approximation (LDA) \cite{dirac1930lda,ceperley1980lda,perdew1981lda} and generalized gradient approximation (GGA) \cite{perdew1996pbe}. More advanced functionals, including meta-GGAs and hybrid functionals, improve the description of semiconductor bandgaps but at substantially higher computational cost, particularly for the large supercells required to model realistic interfaces \cite{weston2018accurate, hinuma2014band}. In addition, the accurate placement of the metal Fermi level, the treatment of interfacial dipoles, and the alignment of electrostatic potentials across the interface introduce further sources of uncertainty. As a result, different computational studies often report widely varying SBHs for the same material systems.

Beyond electronic structure limitations, constructing realistic atomistic models of M-SC interfaces is itself nontrivial. Lattice mismatch, surface terminations, and interfacial reconstructions can strongly influence the local bonding environment and electrostatic potential profile. Consequently, many existing studies focus on a small number of idealized interfaces or employ system-specific modeling choices, limiting the generality and transferability of their conclusions \cite{robertson2013band}. Comprehensive and consistent assessments of Schottky barrier predictions across different materials and computational protocols remain relatively scarce.


Previous first-principles studies have established DFT as a powerful framework for investigating Schottky barriers at metal-semiconductor interfaces, while also highlighting persistent methodological challenges. 

Delaney et al. provided a detailed DFT treatment of Schottky-barrier formation at epitaxial ErAs/GaAs interfaces, using macroscopic averaging of the electrostatic potential to align the metal Fermi level with the semiconductor band edges \cite{delaney2010theoretical}. Their results emphasize that the SBH is highly sensitive to interface orientation, structural details, strain-consistent reference calculations, and DFT bandgap underestimation.

Stengel et al. extended this framework to metal/ferroelectric interfaces and showed that DFT bandgap errors can lead to pathological band alignment, in which the apparent Schottky barrier becomes negative and causes unphysical charge spill-out into the insulator, ultimately making the predicted Schottky barriers unreliable \cite{stengel2011band}.

Later, Wang et al. studied thermionic transport in layered van der Waals heterostructures \cite{wang2016first,wang2018high}. Their work show that predicted barrier heights and currents are highly sensitive to atomistic interface structure, band alignment, and tunneling effects, and they emphasize that standard DFT bandgap errors can significantly affect transport predictions. They further highlight Fermi-level pinning and the limitations of simple band-alignment rules, reinforcing the need for careful interface modeling and beyond-standard-DFT treatments for reliable barrier predictions.

Most recently, Nangoi et al. performed first-principles SBH calculations for Al(111)/Si(111) and \ch{CoSi2}(111)/Si(111) interfaces using the potential-alignment method \cite{nangoi2024first}. Using hybrid-functional bulk references, they showed that SBHs can be strongly structure-dependent in systems such as \ch{CoSi2}/Si, due to more covalent interfacial bonding, while remaining relatively insensitive to interface details in Al/Si. This highlights the strong sensitivity of SBH predictions to interface construction, electrostatic potential alignment, and the choice and accuracy of bulk reference calculations.

\color{black}

Therefore, there is a clear need for a systematic and physically grounded evaluation of the computational strategies used to predict Schottky barriers. In this study, we focus on Si/Metal interfaces as a representative test sample to dissect the role of XC functional choice, reference bulk protocols, structural strain, and relativistic effects in Schottky barrier calculations. By benchmarking multiple computational routes against experimental data and analyzing the sources of error, we aim to define a reliable and transferable methodology for Schottky barrier prediction that can serve as a foundation for future large-scale screening and interface design efforts.

\newpage

\section{Methods}

\subsection{Workflow Overview}
\begin{figure}[h]
\centering
\includegraphics[scale=0.5]{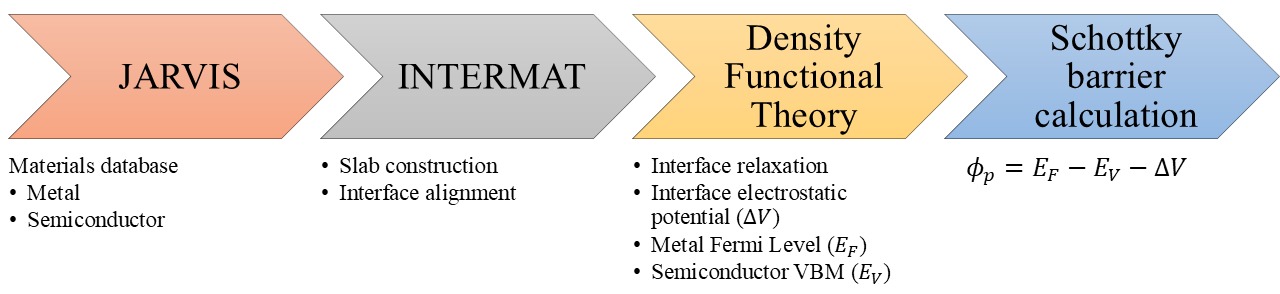}
\caption{Schematic overview of the workflow. Based on the JARVIS-DFT repository, the VBM and $E_F$ are calculated for semiconductors (SC) and metals (M), respectively. Interfaces are generated from the bulk counterparts using the Zur algorithm and ALIGNN-FF. The workflow aims to automate Schottky barrier height calculations.}
\label{fig_workflow}
\end{figure}

Figure~\ref{fig_workflow} presents an overview of the computational procedure used to evaluate Schottky barrier heights at metal-semiconductor interfaces. The framework builds upon the Joint Automated Repository for Various Integrated Simulations density functional theory database (\textbf{JARVIS-DFT}) \cite{choudhary2020jarvis,choudhary2025jarvis}. The approach consists of four main steps: (i) bulk electronic structure calculations to obtain the VBM of the semiconductor and the $E_F$ of the metal, (ii) construction of lattice-matched interface models, (iii) calculation of the electrostatic potential profile across the interface, and (iv) extraction of the Schottky barrier height using the potential alignment formalism.

\subsection{Interface Construction}
To build the M-SC interfaces, we use a combination DFT and machine learning (ML). This effort builds upon our previous work, \textbf{InterMat} \cite{choudhary2024intermat}, a materials interface computational framework designed to predict band offsets of semiconductor interfaces using density functional theory and graph neural networks. In InterMat, we employed a combination of DFT, the \textbf{Zur interface generation algorithm} \cite{zur1984lattice}, and atomistic line graph neural networks (\textbf{ALIGNN}) \cite{choudhary2021alignn} to predict semiconductor-semiconductor (SC-SC) band offsets using both alternate slab junction (ASJ) and independent unit (IU) models.

This work adapts this framework to M-SC systems by generating interfaces using data from the JARVIS-DFT database, applying the Zur algorithm, and pre-relaxing structures with \textbf{ALIGNN-FF}, the Atomistic Line Graph Neural Network force field \cite{choudhary2023alignnff}, enabling efficient screening of large metal-semiconductor combinatorial spaces.

The process begins by selecting surface slabs of metals and semiconductors from the JARVIS-DFT repository and defining the interface parameters. These include maximum lattice mismatch (8 \%), maximum area ($75~\text{\AA}^2$), angular tolerance (1$^\circ$), distance between substrate and film, total cell height, and in-plane displacement interval (0.1~\AA).

Given the importance of the distance between substrate and film, which controls the interfacial bonding interaction and strongly affects the electrostatic potential profile, we determined the optimal separation for each interface using DFT. Calculations were performed by varying the distance from 1.5 to 3.0~\AA\ in steps of 0.1~\AA, selecting the distance that minimized the total energy (see Supplemental Information).

The cell height was optimized with respect to the electrostatic potential by monitoring changes in the planar-averaged potential of the silicon and metal regions in the interface. Calculations were performed by increasing the number of Si bilayers in the interface from 3 to 7 bilayers, while keeping the metal counterpart at a comparable thickness. Changes in the electrostatic potential were found to be minimized for 7 Si bilayers, corresponding to a total cell height of approximately 43.0-43.7~\AA\ for both the metal and semiconductor regions.

Candidate interfaces are constructed by stacking the surface slabs along the $z$-direction. To minimize computational cost, a grid search over in-plane displacements is performed using the ALIGNN-FF machine-learned force field, enabling efficient pre-relaxation of the structures prior to final optimization with DFT.

\subsection{Schottky Barrier Formalism}
\begin{figure}[h]
\centering
\includegraphics[scale=0.7]{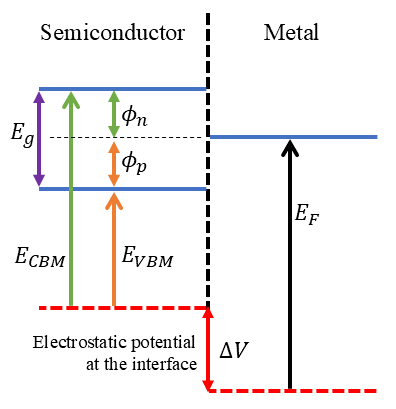}
\caption{Schematic representation of the Schottky barrier at a metal-semiconductor junction.}
\label{fig_schematic}
\end{figure}

The Schottky barrier height is defined as the energy difference between the metal Fermi level and the band edges of the semiconductor. For an ideal, unpinned interface, the hole and electron barriers are given by

\begin{equation}
\Phi_p = E_F - E_{\mathrm{VBM}} - \Delta V,
\end{equation}

\begin{equation}
\Phi_n = E_g - \Phi_p,
\end{equation}

\noindent where $E_F$ is the metal Fermi level, $E_{\mathrm{VBM}}$ is the valence band maximum of the semiconductor, $E_g$ is the semiconductor bandgap, and $\Delta V$ is the electrostatic potential difference across the interface. All quantities are referenced to the average electrostatic potential in their respective bulk regions. The definitions are illustrated schematically in Figure~\ref{fig_schematic}.

The local electrostatic potential $V(x,y,z)$ is extracted from the DFT-calculated Hartree plus ionic potential. To obtain the planar average $\bar{\bar{V}}(z)$ used for band alignment, we perform a two-step averaging.

First, the macroscopic average $\bar{V}(z)$ of the local potential is computed over the cross-sectional area of the interface $S$:
\begin{equation}
\bar{V}(z) = \frac{1}{S} \int_S V(x,y,z)\, dx\,dy.
\end{equation}

Then, the planar average $\bar{\bar{V}}(z)$ is calculated over a repeat length $L$ located in the middle region of the substrate/film:
\begin{equation}
\bar{\bar{V}}(z) = \frac{1}{L} \int_{-L/2}^{L/2} \bar{V}(z+z')\, dz'.
\end{equation}

Finally, the electrostatic potential offset is calculated as
\begin{equation}
\Delta V = \bar{\bar{V}}_\mathrm{SC} - \bar{\bar{V}}_\mathrm{M},
\end{equation}
where a positive $\Delta V$ indicates that the metal $E_F$ lies within the semiconductor $E_g$.

\begin{figure}[t]
\centering
\includegraphics[scale=0.4]{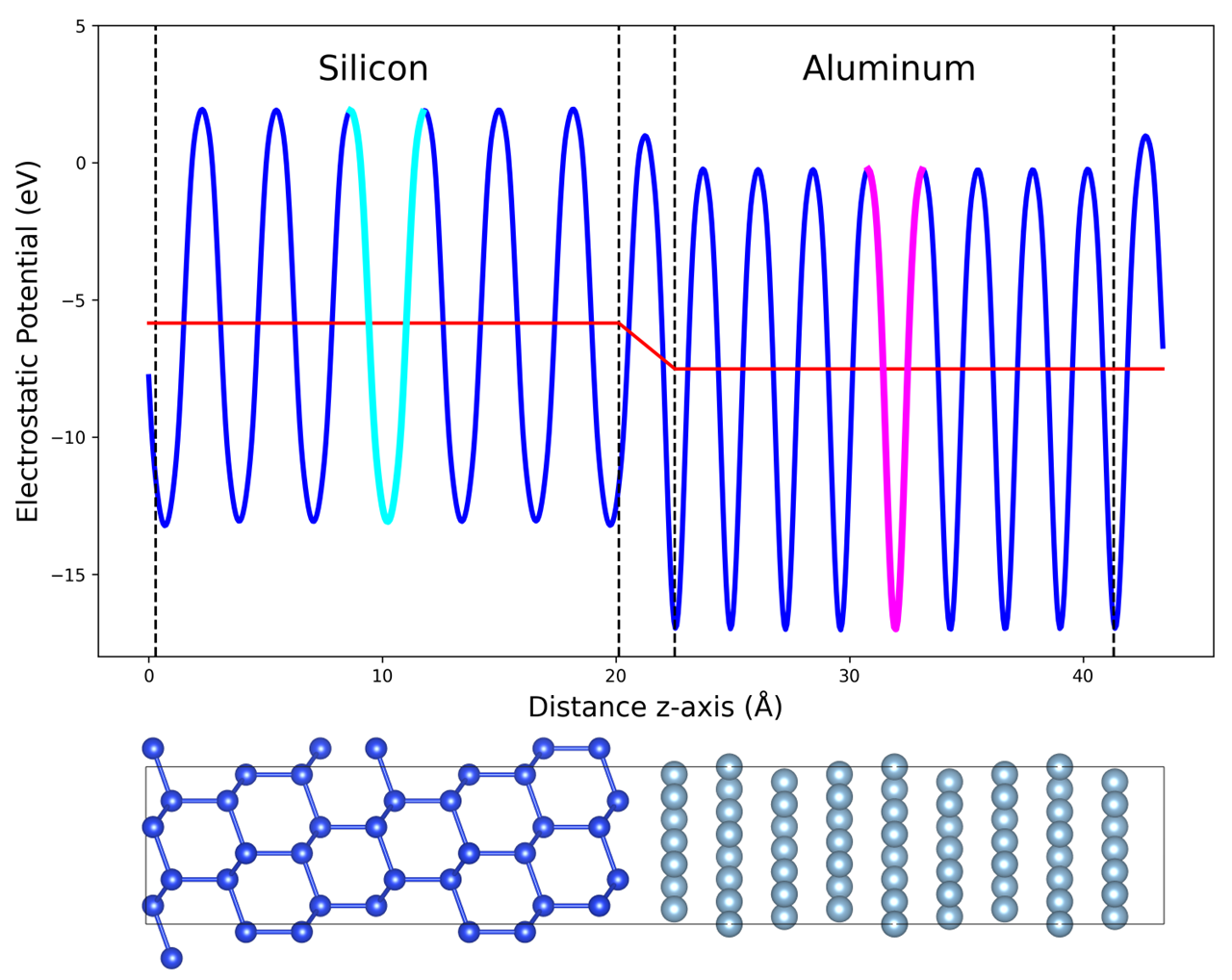}
\caption{(Top) Electrostatic potential of the Si(111)/Al(111) interface. The reference red line represents the average electrostatic potential. Vertical dashed black lines indicate the positions where the Si and Al layers start and end along the z-axis. Cyan and magenta lines denote the repeat unit layers on the left and right sides, respectively. (Bottom) Atomic structure of the Si(111)/Al(111) interface. Si atoms are shown in blue, and Al atoms in gray.}
\label{fig_elec_potential}
\end{figure}

Figure~\ref{fig_elec_potential} (top panel) shows the planar-averaged (red) and macroscopic-averaged (blue) electrostatic potential of the Si(111)/Al(111) interface along the $z$-axis. The vertical dashed lines indicate the boundaries of the interfacial region between the Si and Al layers. The periodic oscillations in the potential reflect the atomic layering within each material, as illustrated in the lower panel of Figure~\ref{fig_elec_potential}, which displays the corresponding atomic structure of the interface.

A key feature of this profile is the electrostatic potential offset across the interface, which corresponds to the dipole term $\Delta V$ in the band alignment equation. The macroscopic average enables identification of bulk-like regions on either side of the interface, where the potential becomes approximately flat. This is essential for accurately referencing the band edges of each material. In the figure, the cyan and magenta curves highlight the bulk-like regions of Si and Al, respectively, and serve as the basis for extracting $\bar{\bar{V}}(z)$ in the calculation of $\Delta V$.

\begin{figure}[h]
\centering
\includegraphics[scale=0.33]{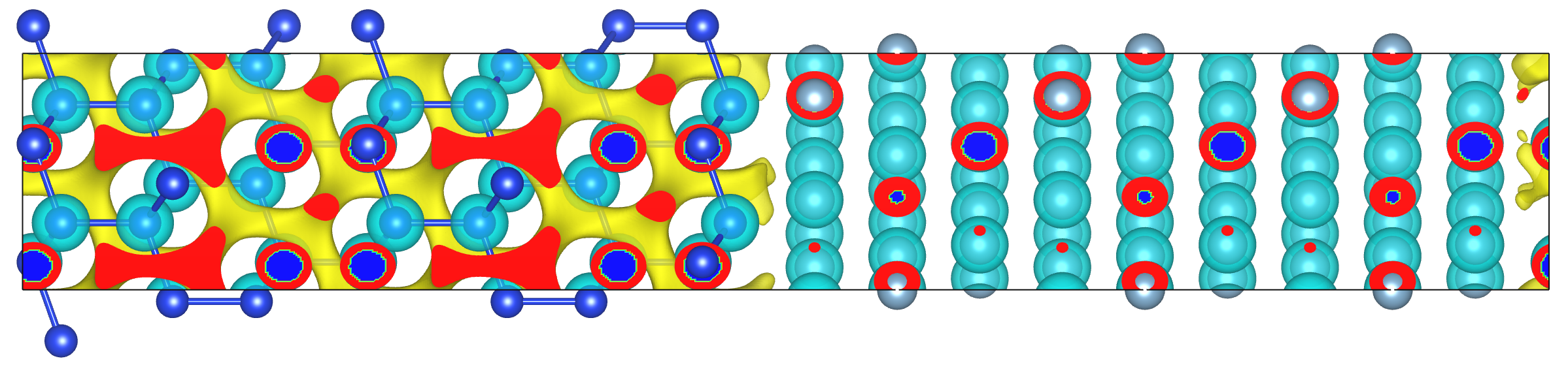}
\caption{Electrostatic potential isosurface of the Si(111)/Al(111) interface rendered in VESTA (Visualization for Electronic and Structural Analysis) at an isovalue of 9.5~eV \cite{Vesta}.}
\label{figure_locpot_Alum}
\end{figure}

To complement the 1D profile, Figure \ref{figure_locpot_Alum} shows a 3D electrostatic isosurface. This visualization captures the spatial variation of electrostatic potential and highlights the interface dipole layer. The isosurface representation is especially useful for detecting localized charge accumulations or discontinuities, which may not be evident in planar-averaged profiles.

\subsection{DFT Details}
All density functional theory calculations were performed using the Vienna \textit{ab initio} Simulation Package (VASP) \cite{vasp1, vasp2}. The projector augmented-wave (PAW) method was employed together with several XC functionals, including Perdew-Burke-Ernzerhof (PBE) \cite{perdew1996pbe}, OptB88vdW (OPT) \cite{klimevs2009optb88vdw}, Strongly Constrained and Appropriately Normed (SCAN) \cite{perdew2015scan}, modified Becke-Johnson (mBJ) \cite{tran2009mbj}, and Heyd-Scuseria-Ernzerhof (HSE, 2006 version) \cite{krukau2006HSE06}.

All calculations used a plane-wave cutoff of 600~eV and Monkhorst-Pack $k$-point meshes that were each converged to better than 1~meV/atom. For all interface supercells, a $7 \times 7 \times 1$ grid was found sufficient to converge both the total energy and the electrostatic potential profile.

Interface relaxations were done at a fixed volume allowing atoms to move (ISIF=2), consistent with the the InterMat framework in which the semiconductor acts as the substrate, while the metal film accommodates most of the lattice mismatch.

\newpage

\section{Results and Discussion}

\subsection{Identification of Stable Interface Configurations}

As a first step, we evaluated the structural stability of the generated metal-semiconductor interfaces in order to identify the most favorable atomic configurations for subsequent Schottky barrier calculations. For each metal (M = Al, Cu, Ag, Au), we constructed Si(111)/M(100), Si(111)/M(110), and Si(111)/M(111) interfaces and assessed their stability using both formation energy and excess energy criteria.

The formation energy of a M-SC interface is defined as the energy required to form the interface from its constituent relaxed bulk phases,
\begin{equation}
E_{\mathrm{formation}} = E_{\mathrm{interface}} - N_{\mathrm{SC}} E_{\mathrm{SC}}^{\mathrm{relaxed}} - N_{\mathrm{M}} E_{\mathrm{M}}^{\mathrm{relaxed}},
\end{equation}
where $E_{\mathrm{interface}}$ is the total energy of the interface supercell, $E_{\mathrm{SC}}^{\mathrm{relaxed}}$ and $E_{\mathrm{M}}^{\mathrm{relaxed}}$ are the energies per atom of the relaxed bulk semiconductor and metal, respectively, and $N_{\mathrm{SC}}$ and $N_{\mathrm{M}}$ are the corresponding numbers of atoms in the interface cell.


To account for the presence of two interfaces in the periodic supercell and to enable comparison across different interface sizes, we further compute the work of adhesion ($W_{\mathrm{ad}}$) as \cite{book2013surface}:
\begin{equation}
W_{\mathrm{ad}} = \frac{N_{\mathrm{SC}} E_{\mathrm{SC}}^{\mathrm{relaxed}} + N_{\mathrm{M}} E_{\mathrm{M}}^{\mathrm{relaxed}} - E_{\mathrm{interface}}}{2A},
\end{equation}
where $A$ is the interfacial area. The work of adhesion is the reversible work required to separate two contacting phases from their equilibrium interfacial distance to infinite separation. Thus, larger $W_{\mathrm{ad}}$ values correspond to stronger interfacial bonding and greater interface stability. The calculated $W_{\mathrm{ad}}$ includes contributions from interfacial bonding and residual elastic strain associated with lattice matching, and therefore serves as a practical measure of interface stability.

\color{black}

All stability calculations were performed using self-consistent PBE calculations for computational efficiency. For all metals in this study, the Si(111)/M(111) interface was found to be the most stable configuration, exhibiting the lowest formation and excess energies compared to the (100) and (110) terminations. Detailed numerical values for all configurations are reported in Table~S1 of the Supporting Information.

Based on these results, all subsequent electronic structure and Schottky barrier calculations in this work focus exclusively on the Si(111)/M(111) interfaces.

\subsection{Schottky Barrier Calculations}

\begin{figure}[h]
\centering
\includegraphics[scale=0.55]{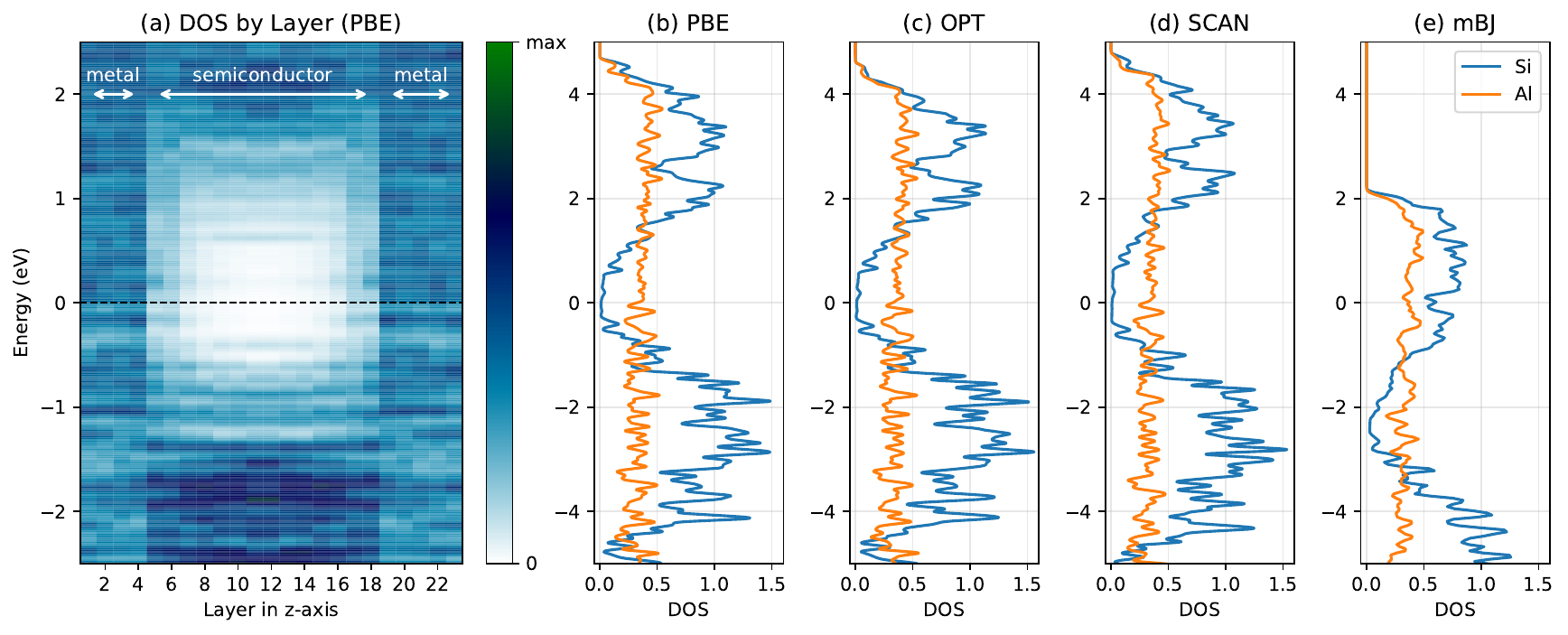}
\caption{Local density of states (LDOS) of the Al/Si interface computed with different XC functionals: (a) layer-resolved LDOS along the interface normal (z-axis) using PBE; (b-e) LDOS for the central layers of the Al and Si slabs computed with (b) PBE, (c) OPT, (d) SCAN, and (e) mBJ. In all panels, energies are referenced to the Fermi level ($E_F = 0$).}
\label{figure_DOS_alum}
\end{figure}

Figure \ref{figure_DOS_alum}(a) shows the layer-resolved Local density of states (LDOS) of the Al/Si interface computed with PBE. The low-LDOS (white) region in the middle of the Si slab corresponds to the Si bandgap. However, a strictly vanishing LDOS is not recovered. As seen in Figures \ref{figure_DOS_alum}(b-e), the Si layers progressively develop a gap-like feature around the Fermi level, but residual spectral weight persists in the nominal gap. This behavior is mainly attributed to metal-induced gap states (MIGS) and Al-Si wave function hybridization at the interface, and it can be enhanced by the finite smearing used in the DOS calculation. Notably, the gap-like Si region improves from PBE to OPT to SCAN. In contrast, mBJ deviates from this trend, showing an excessive upward shift of the Fermi level that yields an artificially metallic LDOS in the middle of the Si slab.
\color{black}

\newpage

\begin{table*}[h]
\centering
\caption{Schottky barrier heights (eV) for Si(111)/M(111) interfaces computed using different XC functionals and reference protocols.
Procedure A: relaxed bulk references;
Procedure B: relaxed bulk references with SOC;
Procedure C: strained, vacuum-free bulk references consistent with the interface geometry.
Experimental values extracted from Ref. \cite{smith1971schottky}  are shown in the last column. Mean Absolute Error (MAE) is reported for each XC functional vs experiment. Negative SB values are highlighted in bold red and the best MAE in each procedure is bolded.}
\label{table_results}
\renewcommand{\arraystretch}{1.2}
\setlength{\tabcolsep}{6pt}
\begin{tabular}{l c c c c c c c c}
\hline
\rowcolor{gray!15}
\multicolumn{9}{l}{\textbf{Procedure A (relaxed)}}\\
\rowcolor{gray!15}
Metal & PBE & OPT & SCAN & mBJ & HSE+PBE & HSE+OPT & HSE+SCAN & Expt \\
Al  & 0.58 & 0.71 & 0.88 & 0.56 & 0.75 & 0.94 & 0.91 & 0.58 \\
Cu  & \negSB{-0.27} & \negSB{-0.09} & 0.15 & 0.71 & \negSB{-0.10} & 0.13 & \negSB{-0.01} & 0.46 \\
Ag  & 0.01 & \negSB{-0.08} & 0.17 & 1.10 & 0.18 & \negSB{-0.05} & \negSB{-0.10} & 0.55 \\
Au  & \negSB{-0.12} & \negSB{-0.12} & 0.11 & 0.76 & 0.05 & 0.10 & \negSB{-0.15} & 0.34 \\
MAE & 0.43 & 0.44 & \textbf{0.31} & \textbf{0.31} & 0.35 & 0.38 & 0.49 & -- \\
\hline

\rowcolor{gray!15}
\multicolumn{9}{l}{\textbf{Procedure B (relaxed+SOC)}}\\
\rowcolor{gray!15}
Metal & PBE & OPT & SCAN & mBJ & HSE+PBE & HSE+OPT & HSE+SCAN & Expt \\
Al  & 0.56 & 0.69 & 0.86 & 0.55 & 0.74 & 0.94 & 0.91 & 0.58 \\
Cu  & \negSB{-0.29} & \negSB{-0.10} & 0.16 & 0.70 & \negSB{-0.14} & 0.08 & \negSB{-0.05} & 0.46 \\
Ag  & 0.01 & \negSB{-0.08} & 0.17 & 1.09 & 0.18 & \negSB{-0.06} & \negSB{-0.10} & 0.55 \\
Au  & 0.04 & 0.03 & 0.26 & 0.76 & 0.20 & 0.25 & \negSB{-0.01} & 0.34 \\
MAE & 0.40 & 0.40 & \textbf{0.26} & 0.31 & 0.32 & 0.36 & 0.46 & -- \\

\hline
\rowcolor{gray!15}
\multicolumn{9}{l}{\textbf{Procedure C (strained)}}\\
\rowcolor{gray!15}
Metal & PBE & OPT & SCAN & mBJ & HSE+PBE & HSE+OPT & HSE+SCAN & Expt \\
Al  & 0.00 & 0.28 & 0.36 & 0.12 & 0.36 & 0.60 & 0.57 & 0.58 \\
Cu  & 0.03 & 0.19 & \negSB{-0.01} & 0.92 & 0.39 & 0.61 & 0.46 & 0.46 \\
Ag  & 0.20 & \negSB{-0.06} & \negSB{-0.04} & 0.73 & 0.56 & 0.33 & 0.28 & 0.55 \\
Au  & 0.13 & 0.18 & \negSB{-0.22} & 1.05 & 0.49 & 0.54 & 0.28 & 0.34 \\
MAE & 0.39 & 0.34 & 0.46 & 0.45 & 0.11 & 0.15 & \textbf{0.09} & -- \\
\hline
\end{tabular}
\end{table*}

As shown in Table~\ref{table_results}, we evaluate the performance of several XC approaches for predicting SBHs at Si/Metal interfaces, including semilocal (PBE, OPT), meta-GGA (SCAN), modified Becke-Johnson (mBJ), and mixed hybrid-semilocal schemes (HSE+PBE, HSE+OPT, HSE+SCAN). For the mBJ calculations, we used a fixed setting of $\beta=0.9797$ with the default $\alpha$ parameter for the interface, bulk-Si, and bulk-metal calculations. The $\beta$ value was chosen to reproduce the experimental Si bandgap ($E_g=1.17$~eV), and the mBJ calculations were performed on structures taken from previously relaxed OPT calculations. The mixed hybrid-semilocal schemes denote calculations in which the interface electrostatic lineup $\Delta V$ is obtained from a semilocal XC functional (PBE, OPT, or SCAN), while the bulk reference quantities are computed using HSE. In these HSE calculations, the Si exact-exchange fraction was set to $\alpha=0.23$ to reproduce $E_g=1.17$~eV, whereas the metals were evaluated with $\alpha=0$ to avoid introducing artificial exchange effects. No explicit HSE interface calculations were performed due to their excessive computational cost. In addition, we assess three distinct reference protocols for extracting bulk quantities entering the potential-alignment formalism: (A) relaxed bulk silicon and metal calculations, (B) relaxed bulk silicon and metal calculations including spin-orbit coupling (SOC), and (C) strained, vacuum-free bulk silicon and metal calculations consistent with the interface geometry. 

\subsubsection{Relaxed bulk reference calculations}

Using relaxed bulk references \textbf{(Procedure A)}, PBE and OPT systematically underestimate SBHs and frequently predict negative barriers for Cu-, Ag-, and Au-based interfaces, indicating unphysical metallic alignment. SCAN and mBJ yield uniformly positive SBHs, but exhibit distinct biases: SCAN tends to underestimate barriers, while mBJ significantly overestimates SBHs for noble metals due to an unphysical upward shift of the metal $E_F$. Overall, SCAN provides the most balanced performance within Procedure A, albeit with moderate mean absolute errors.

\subsubsection{Including spin-orbit coupling in bulk references}

Including SOC in relaxed bulk reference calculations \textbf{(Procedure B)} primarily affects heavy metals, most notably Au. While SOC improves the SBH prediction for Au in selected approaches, it does not systematically resolve negative barriers or alignment inconsistencies across all methods. In particular, PBE, OPT, and mixed approaches that already suffer from unphysical SBHs remain problematic. SCAN with SOC shows a modest reduction in error relative to Procedure A, whereas mBJ retains its overestimation trend. These results indicate that SOC provides a secondary correction that is insufficient on its own to ensure reliable SBH predictions.

\subsubsection{Strained, vacuum-free bulk references}

The most significant improvement in both physical consistency and quantitative accuracy is achieved using strained, vacuum-free bulk silicon and metal calculations consistent with the interface geometry \textbf{(Procedure C)}. Under this protocol, several approaches that previously produced negative or scattered SBHs yield uniformly positive and substantially more accurate results. This highlights the critical importance of using reference bulk calculations that preserve the strain state and periodicity imposed by the interface.

Within Procedure C, mixed hybrid-semilocal approaches clearly outperform purely semilocal or meta-GGA functionals. In particular, HSE+SCAN provides the lowest overall error, closely reproducing experimental SBHs across all interfaces. HSE+PBE and HSE+OPT achieve comparable accuracy with slightly higher errors, but at reduced computational cost. In contrast, PBE and mBJ, while yielding positive barriers under this protocol, remain less accurate overall.

Taken together, these results demonstrate that structural and electrostatic consistency between interface and bulk reference calculations is the dominant factor governing SBH accuracy, outweighing the isolated impact of SOC or XC functional choice alone. While HSE+SCAN combined with Procedure C provides the highest accuracy, the computational cost of SCAN-based interface calculations limits its scalability. As a practical compromise for high-throughput screening, we identify HSE+PBE with strained, vacuum-free bulk references (Procedure C) as an optimal choice, delivering near-benchmark accuracy with substantially reduced cost.

SOC may be applied selectively for heavy-metal contacts such as Au, where relativistic effects are expected to be significant, but is not required for the default high-throughput SchottkyMat workflow.

\section{Conclusion}
In this work, we systematically assessed the reliability of density functional theory-based approaches for predicting SBHs at Si/Metal interfaces using the potential alignment method. By benchmarking multiple XC functionals and reference protocols against experimental data for Si(111)/Metal (Al, Cu, Ag, and Au) interfaces, we identified the key factors governing both physical consistency and quantitative accuracy.

We show that commonly used semilocal and nonlocal functionals such as PBE and OPT frequently yield unphysical (negative) SBHs, rendering them unsuitable for predictive screening. Meta-GGA SCAN and mBJ consistently produce positive barriers, but with distinct and systematic biases: SCAN provides balanced but underestimated SBHs, while mBJ strongly overestimates barriers for metals due to an artificial upward shift of the metal Fermi level. Inclusion of spin-orbit coupling improves agreement for heavy metals such as Au, but does not resolve fundamental alignment errors and is therefore a secondary correction.

The most significant improvement arises from enforcing structural and electrostatic consistency between interface and bulk reference calculations. When bulk silicon and metal references are computed under the same strain conditions as the interface, mixed hybrid-semilocal approaches yield uniformly positive SBHs with substantially reduced error. Among these, HSE+SCAN achieves the highest accuracy, while HSE+PBE provides comparable performance at significantly lower computational cost, making it the preferred choice for high-throughput applications.

Overall, our results demonstrate that accurate Schottky barrier prediction depends more critically on consistent reference protocols than on increasing XC sophistication alone. The proposed framework offers a robust and transferable strategy for large-scale screening of metal-semiconductor interfaces and can be readily extended to other semiconductors, metals, and interface orientations. This work lays the foundation for data-driven discovery and optimization of contact materials in next-generation electronic and optoelectronic devices.

\section*{Acknowledgments}
Official contribution of the National Institute of Standards and Technology (NIST); not subject to copyright in the United States.
All authors thank NIST for funding, computational, and data-management resources. This work was supported by the Creating Helpful Incentives to Produce Semiconductors (CHIPS) Metrology Program, part of CHIPS for America at NIST, U.S. Department of Commerce. Certain commercial equipment, instruments, software, or materials are identified in this paper to adequately specify the procedures used. Such identification is not intended to imply recommendation or endorsement by NIST, nor is it intended to imply that the materials or equipment identified are necessarily the best available for the purpose. The authors acknowledge Kevin Garrity (NIST) for fruitful discussions regarding the manuscript and the software package.

\bibliography{refs}

\end{document}